\documentclass{article}
\usepackage{graphicx} % Required for inserting images
\usepackage[utf8]{inputenc}

\usepackage{latexsym}
\usepackage{verbatim}
\usepackage[a4paper, total={6.5in, 9in}]{geometry}
\usepackage{silence}
\usepackage{amsthm,amsmath,amsfonts,amssymb,verbatim, color}
\usepackage{graphicx}
\usepackage[percent]{overpic}
\usepackage{bm}
\usepackage{braket}
\usepackage{float}
\usepackage{mathtools}
\usepackage{appendix}
\usepackage[colorlinks=true,citecolor=blue,linkcolor=black,urlcolor=blue]{hyperref}
\usepackage{graphics} % Testing...
\graphicspath{ {figures/} }
\usepackage[caption=false]{subfig}

\usepackage[mathscr]{euscript}
\usepackage[export]{adjustbox}
\usepackage[bottom]{footmisc}
\usepackage{textcomp}
\usepackage{braket}
\title{\bf Quantum error correction beyond the toric code: dynamical systems meet encoding}
\author{$^{1}$Garima Rajpoot, $^{1,2}$Komal Kumari, and $^{1,2}$Sudhir Ranjan Jain$^*$ \\ $^{1}${\small{\it Theoretical Nuclear Physics and Quantum Computing Section}}\\ {\small{\it Nuclear Physics Division, Bhabha Atomic Research Centre, Mumbai 400085, India}}\\ $^{2}${\small{\it Homi Bhabha National Institute, Training School Complex, Anushakti Nagar, Mumbai 400094, India}}\\ $^*$srjain@cbs.ac.in
}
\date{February 2023}

\begin{document}

\maketitle

\begin{abstract}
We construct surface codes corresponding to genus greater than one in the context of quantum error correction. The architecture is inspired by the topology of invariant integral surfaces of certain non-integrable classical billiards. Corresponding to the fundamental domains of rhombus and square torus billiard, surface codes of genus two and five are presented here. There is significant improvement in encoding rates and code distance, in addition to immunity against noise.  
\end{abstract}

\section{From geometry to encoding}

Geometrical representations of algebraic and arithmetic relations \cite{kvant,weissman}, and, algebraic representations of geometrical patterns \cite{aop2014, aop2016, rmp2017} are both fascinating themes. In their turns, they have led to a deep understanding in physics and mathematics \cite{nakahara}. A one-to-one correspondence between Lie groups and reflection groups whose fundamental regions are simplexes in Euclidean space has been beautifully illustrated in \cite{coxeter, weyl1926nachtrag,cartan1927geometrie}. These fundamental regions generate tori for ``unit shapes" like a square, equilateral triangle, right isosceles triangle, or a hemi-equilateral triangle \cite{rmp2017}. Here we bring out an application of geometry of regular polytopes \cite{coxeter} to encoding theory in the context of quantum information. 

The dynamical systems which are most relevant to the present theme are planar billiards wherein a particle moves freely inside a two-dimensional  enclosure, reflecting from the boundary in accordance to the Snell's law of reflection. According to the Liouville-Arnol'd theorem \cite{arnold}, for a system with $f$ degrees of freedom, if there are $f$ functionally independent invariants which are in involution, the (invariant) surface on which the trajectory of the system resides is topologically equivalent to an $f$-torus. Another condition stipulated for the applicability of the Liouville-Arnol'd theorem is that the vector fields in phase space must be smooth everywhere. The integrability of such systems is a fragile property, so much so that even if the vector fields become singular at points of measure zero, the system loses integrability \cite{jain1992}. Perhaps the simplest example is when the shape of the enclosure is a square or a rectangle, explained later in some detail, where the invariant surface is a torus. However, an interesting situation arises by deforming the square to a rhombus with an acute angle $\pi /n$. The vector fields in phase space become singular at a set of points of measure zero. Corresponding invariant surface is topologically equivalent to a sphere with few handles, the number of handles is related to $n$. In this work, instead of a lattice of spins, we employ the lattice constructed by stacking fundamental domains in a plane. On this lattice, we show how to place qubits and set up a stabilizer code.         

Somewhat unrelated but of great significance, a connection between billiard and computation was first realized by Fredkin and Toffoli \cite{toffoli}. Although it gave us the Toffoli gate, the connection between topology of invariant surfaces in billiards and surface codes was not  relevant for them and has been brought out recently \cite{krj}.

\section{Genus-2 code}
Computation requires scalability of logical qubits on planar chips. One way to achieve this is to use ``unit shapes" which can fill the plane on successive reflections to encode the information on a surface. Our aim is to make use of the fundamental domains of certain geometrical structures such as squares and rhombi, which upon successive reflections, fill the whole plane while maintaining the planarity of the surface. This suitable arrangement allows one to make changes anywhere else in the circuit by only locally changing parameters, inadvertently leading to scalability. For example, if we consider a square tile, upon successive reflections about its sides, four copies form a unit of tessellation - the fundamental domain, identifying the pairs of parallel edges gives a torus, which is characterized by a topological invariant, the genus being equal to one. Thus, the surface code corresponds to tori, and hence makes the well-known ``toric code" \cite{Kitaev}. The fundamental domain of a $\pi/3$-rhombus is another such structure, genus equal to two, that can be tessellated on the whole surface. Here, we use this to design a new code on a surface of genus two.

\subsection{``Tessellation" with \texorpdfstring{$\pi/3$}{Lg}-rhombus}
\begin{figure}
    \begin{center}
    \includegraphics[width=0.95\textwidth]{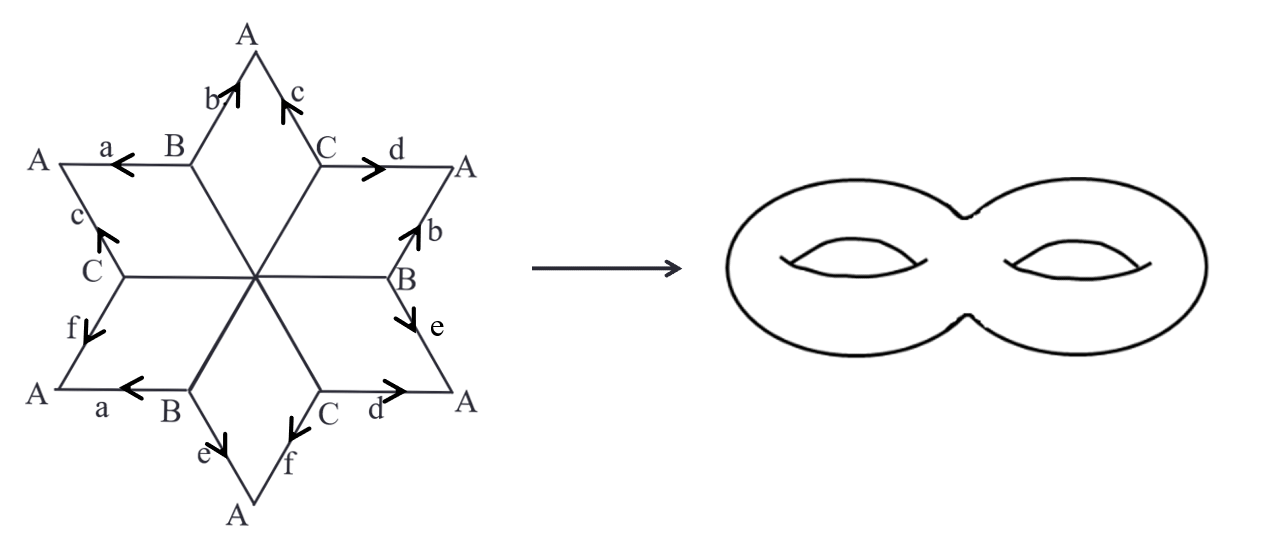}
    \caption{\small{The fundamental domain of a $\pi/3$-rhombus is formed by using six copies of the rhombus, thus constructing an invariant surface. By identifying the edges labelled by the same arrows, the shape shown folds into a surface topologically equivalent to a sphere with two handles (genus, g = 2).}}
    \label{fig:fd}
    \end{center}
\end{figure}

\begin{figure}
    \begin{center}
    \includegraphics[width=0.65\textwidth]{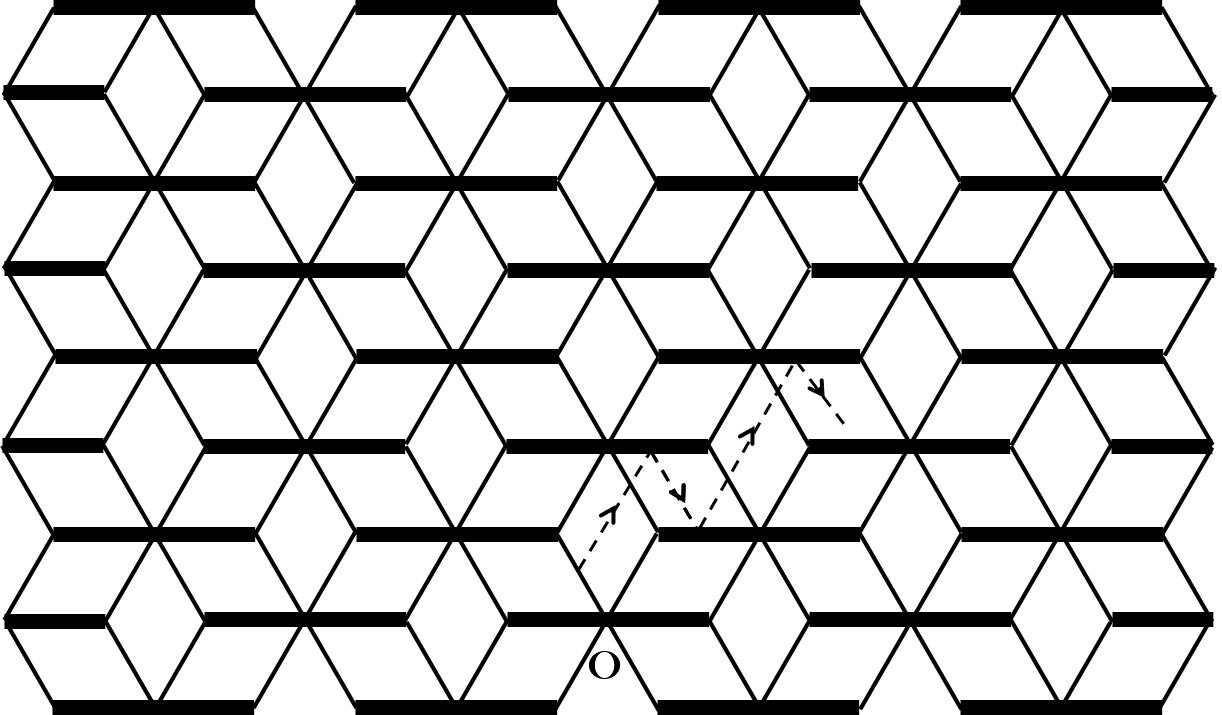}
    \caption{\small{A plane can be filled by successively reflecting a $\pi/3$-rhombus about its sides. The arrangement of the rhombi is shown here with bold segments representing the regions visited twice. These bold segments are the branch cuts of the structure. Upon identification of corresponding sides, one obtains the surface of a sphere with two handles (see Fig. \ref{fig:fd}). To compare, with square as a basic unit, one obtains the surface of a torus \cite{rmp2017}.}}
    \label{fig:domain}
    \end{center}
\end{figure}

We introduce a new surface code using the fundamental domain equivalent to a genus two surface (Fig. \ref{fig:fd}), constructed by stitching six copies of $\pi/3$-rhombus. Upon identification of edges as shown in Fig. \ref{fig:fd}, it creates a ``double-torus" \cite{eckhardt1984analytically,richens1981pseudointegrable,rmp2017}, which is equivalent to a sphere with two handles. This can be tessellated over the whole plane as shown in Fig. \ref{fig:domain}. Hence, encryption on this surface is termed as ``Genus two code" or ``Double-toric code". As per Kitaev's idea, whereby increasing the genus will give a higher encryption, the double-toric code helped achieve a significantly higher encryption rate as compared to the surface code.

\subsection{Encoding on a plane}
\begin{figure*}[t!]
    \centering
    \subfloat[]{\includegraphics[width=0.45\textwidth]{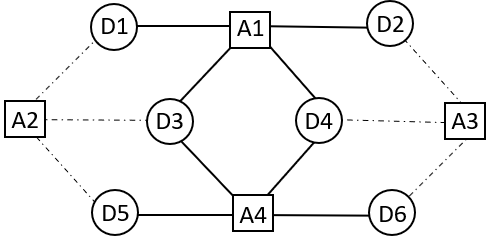}}\hspace{0.8mm}\\
    \subfloat[]{\includegraphics[width=0.65\textwidth]{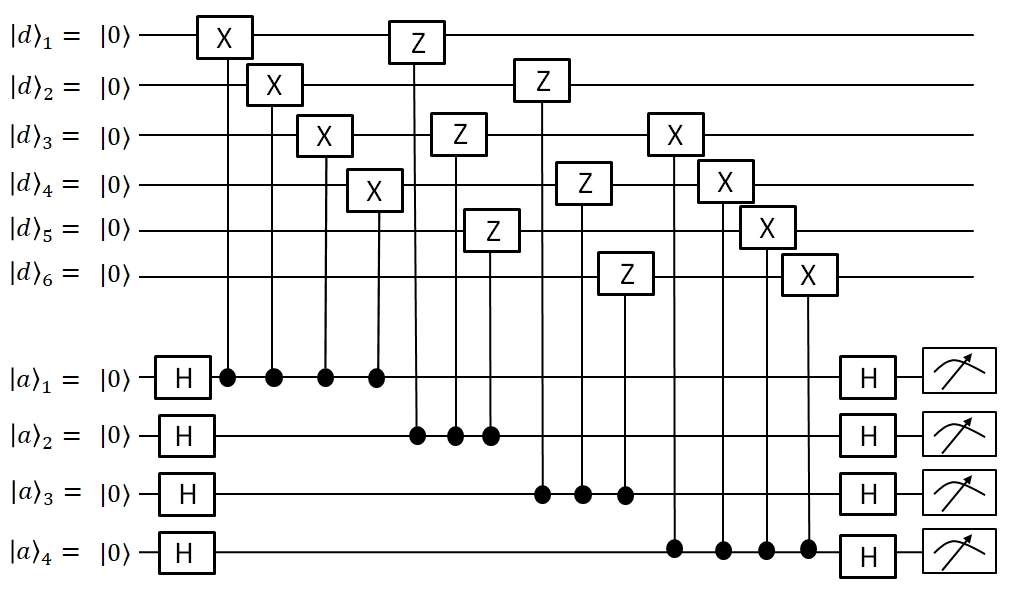}} 
    \caption{\small{(a) Using the fundamental domain of $\pi/3$-rhombus, which is equivalent to a genus two surface, the unit structure of the code is constructed. The data qubits are represented by $D$ and ancillary qubits are represented by $A$. Bold (dashed) lines represent the control $X(Z)$ operation. In (b), the circuit diagram of the unit structure (a) to encode the logical state $\ket{0}_L$  is shown. The $\ket{d}_i(\ket{a}_j)$ represents the initial state of the $D_i(A_j)$ data(ancilla) qubit.}}
\label{fig:G2}
\end{figure*}

Let us start with a unit structure of the genus two code - constructed by using $n=6$ data qubits (represented by circles) and $m=4$ ancilla qubits (represented by squares), shown in Fig. \ref{fig:G2}.
The bold and dashed lines represent the control-$X$ and control-$Z$ operations, respectively from the ancilla qubit to the data qubits. Stabilizers are the operators which belong to the Pauli group and preserve the logical state, i.e. if the logical state is $\ket{\Psi}_L$, then $P_i\ket{\Psi}_L=(+1)\ket{\Psi}_L$. The set of stabilisers for this code structure is $P=\{X_1X_2X_3X_4$, $X_3X_4X_5X_6$, $Z_1Z_3Z_5$, $Z_2Z_4Z_6\}$. These four elements of the stabilizer set are the generators of the stabilizer group ${\mathcal S}$. For this encoded logical qubit, the logical state $\ket{0}_L$ is \cite{Gottesman}:
\begin{alignat}{1}\label{eq:zerol_g2}
    \ket{0}_L&=\frac{1}{\mathcal{N}}\prod_{P_i\in\langle P\rangle}(I^{\otimes n}+P_i)\ket{0^{\otimes n}}\nonumber\\
    &=\frac{1}{\mathcal{N}}(I^{\otimes 6}+X_1X_2X_3X_4)(I^{\otimes 6}+X_3X_4X_5X_6)(I^{\otimes 6}+Z_1Z_3Z_5)(I^{\otimes 6}+Z_2Z_4Z_6)|0^{\otimes 6}\rangle\nonumber\\
    &=\frac{1}{\mathcal{N}}(\ket{000000}+\ket{001111}+\ket{111100}+\ket{110011}),
\end{alignat}

where ${\mathcal N}$ is the normalization factor. The circuit for this encryption is shown in Fig. \ref{fig:G2} (b). All the stabilizers commute with each other ($[P_i,P_j]=0$ $\forall$ $i,j$). To construct logical state $\ket{1}_L$, we have to look for analogous Pauli sigma pairs of logical operators $\{\Bar{X}_i,\Bar{Z}_i\}$, that (i) commute with each of the stabilizers $P_j$ ($[\Bar{X}_i,P_j]=0=[\Bar{Z}_i,P_j]$ $\forall$ $i,j$) and (ii) pairwise anti-commute with each other ($\{\Bar{X}_i,\Bar{Z}_i\}=0$ and $[\Bar{X}_i,\Bar{Z}_j]=0$ $\forall$ $i\neq j$). 

To find the logical operators, first we have to identify the edges to specify the boundaries. The filling of plane using $\pi/3$-rhombus, forms periodically arranged branch-cuts, which help identify the boundaries. On these boundaries, the control-$X$ (bold lines) and control-$Z$ (dashed lines) are arranged alternately. We define a path, between the boundaries, by connecting a data qubit vertex of a rhombus to another data qubit vertex of a corresponding copy with respect to the fundamental domain of the rhombus. Two sets of six paths are found which form the logical $X$ operator ($\bar{X}$) and logical $Z$ operator($\bar{Z}$). Thus we found two pairs of logical operators, which satisfy the above conditions $\{\bar{X}_1=X_1X_3$, $\bar{Z}_1=Z_1Z_4Z_6\}$ and $\{\bar{X}_2=X_4X_6$, $\bar{Z}_2=Z_2Z_4Z_5\}$. The minimum weight of error $E=E_a^\dagger E_b$ violating the Knill-Laflamme conditions \cite{Gottesman} was found to be $2$. Thus it is a $[[6,2,2]]$ code. The encoding rate, or the ratio of the number of logical qubits to the number to data qubits for this code structure is $1/3$.

To increase the code distance and the encoding rate of the double-toric code, we can stack a unit of this code (Fig. \ref{fig:G2}) vertically as well as horizontally. Reflecting the unit in equal number of vertical and horizontal directions, arranges the unit structures in equal number of rows and columns. To construct the code with $p^2$ number of unit structures, the number of rows and columns will be $p$, the the number of required data qubits is $n=2p(2p+1)$, number of required ancilla qubits is $m=2p(p+1)$, number of logical qubit is $k=2p^2$ and the code distance is $d=\lfloor\frac{p+2}{2}\rfloor+1$, where $\lfloor \cdot\rfloor$ is the floor function. So the general form of the code is $\left[\left[2p(2p+1), 2p^2,\lfloor \frac{p+2}{2} \rfloor+1\right]\right]$. The encoding rate of this code is $k/n=p/(2p+1)$. For $p\to\infty$, the encoding rate is $1/2$.

\subsection{Comparison of code distance in toric and genus-2 codes}

\begin{figure*}[ht!]    
    \centering
    \subfloat[]{\includegraphics[width=0.4\textwidth]{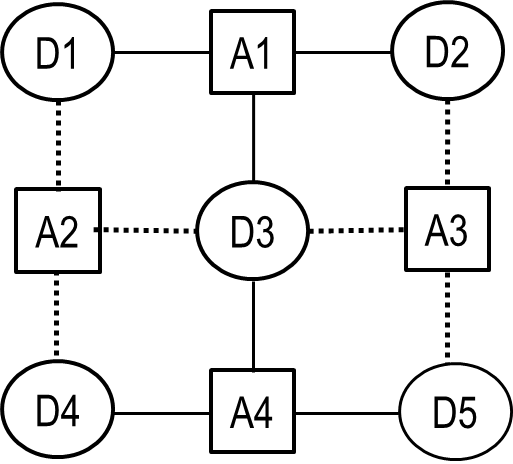}}\hspace{15mm}
    %\subfloat[]{\includegraphics[width=0.35\textwidth]{square_reflections.png}}  
    %\subfloat[]{\includegraphics[width=0.35\textwidth]{ent_al_1.7_x1-x7_0.4.png}}
    %\subfigure[]{\includegraphics[width=0.24\textwidth]{monalisa.jpg}}
    \caption{In [[5,1,2]] surface code shown here, $Ai$'s represent the ancillae and $Di$'s represent  the data qubits. Bold (dotted) lines depict Control-$X$ ($Z$) operations.}
\label{512code}
\end{figure*}

In the [[5,1,2]] code shown in Fig. \ref{512code}, the code distance is $2$. Let us try to make a logical operator of weight 3. The paths $D1-A1-D3-A4-D5$ and $D2-A3-D3-A2-D4$ provide such a pair of logical operator $\langle\bar{X}=X_2X_3X_4, \bar{Z}=Z_1Z_3Z_5\rangle$. Both the operators commute with all the stabilizers of the $[[5,1,2]]$ code and anticommute with each other. In this way we achieved a pair of logical operators of weight $3$ and so the code distance could be 3 making it a [[5,1,3]] code instead. But for the states corresponding to these operators, the minimum weight of error for which  Knill-Laflamme conditions do not hold is $d=2$, indicating that this has to be a distance $2$ code, hence the code is $[[5,1,2]]$. This is well-expected. 

It is important to note that we could have found all logical operators of weight $2$, while maintaining the code distance two - $\{X_1X_3, Z_1Z_2\}$ and $\{X_4X_6, Z_5Z_6\}$. In this case also, the minimum weight of errors for which the Knill-Laflamme conditions do not hold is two. So we could have chosen either set of logical operators. But it is our aim to maximize the code distance using the reflection property of the structure. This makes the [[$2p(2p+1),2p^2,\lfloor\frac{p+2}{2}\rfloor+1$]] code more suitable for achieving higher encryption rates and distances than a [[$2p(2p+1),2p^2,2$]] code.

\begin{figure*}[t!]
    \centering
    \subfloat[]{\includegraphics[width=0.45\textwidth]{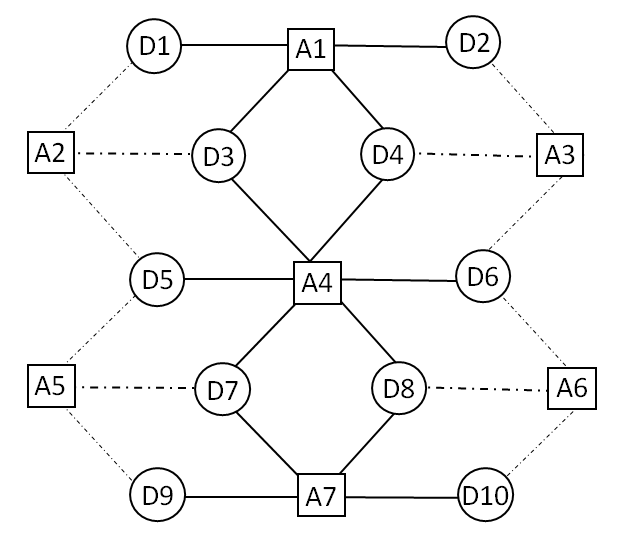}}\hspace{0.8mm} 
    \caption{\small{The unit in Fig. \ref{fig:G2} has been copied vertically. This increases the number of encryptions by only one, giving a total of three logical operators, but the distance has now become three, providing us the possibility of single-qubit error correction.}}
\label{fig:10_3_3}
\end{figure*}

Consider now another unit stacked vertically on the single unit as shown in Fig. \ref{fig:10_3_3}. Here, the number of physical qubits is $n=10$, while the number of ancilla qubits is $m=7$. The stabilizers for this code are, $P=\{X_1X_2X_3X_4$, $X_3X_4X_5X_6X_7X_8$, $X_7X_8X_9X_{10}$, $Z_1Z_3Z_5$, $Z_2Z_4Z_6$, $Z_5Z_7Z_9$, $Z_6Z_8Z_{10}\}$. Following the arguments presented above for identifying paths between boundaries, we obtain  $\bar{X}$ and $\bar{Z}$; the complete set of logical operators commuting with the stabilizers and anti-commuting pairwise is thus 
(i) \{$\bar{X}_1=X_2X_6X_8$, $\bar{Z}_1=Z_1Z_4Z_8Z_9$\}, 
(ii) \{$\bar{X}_2=X_2X_6X_{10}$, $\bar{Z}_2=Z_5Z_7Z_{10}$\}, 
(iii) \{$\bar{X}_3=X_4X_6X_8$, $\bar{Z}_3=Z_2Z_3Z_6$\}. 
The Knill-Laflamme conditions are violated for a weight of error three, giving the code distance three. However, we can again find logical operators of weight two - $\{X_1X_3,Z_1Z_2\}$, $\{X_3X_5X_7,Z_5Z_6\}$ and $\{X_7X_9,Z_9Z_{10}\}$. This should give a distance of two which is also verified using the Knill-Laflamme conditions. Since both the cases are valid, we choose to use the one in which the distance is maximum without violating the stabilizer algebra. 

\section{Genus-5 code}
The motivation to this code stems from another dynamical system, the square torus billiard where the integrable dynamics of a square billiard is interrupted by a square shaped scatterer \cite{aa}. Following the association discussed above for genus 2, we construct a code with this dynamical system in mind. 

\subsection{Square torus billiard}
\begin{figure}
    \begin{center}
    \includegraphics[width=0.35\textwidth]{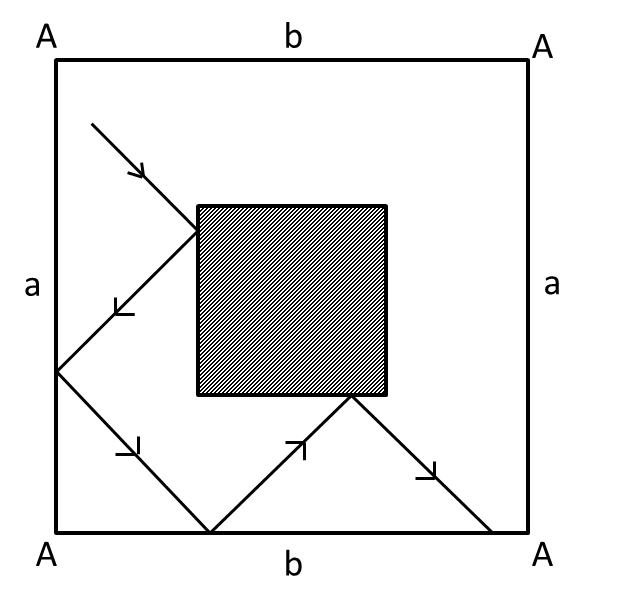}
    \caption{\small{The arrow guiding us here depicts the motion of a point particle moving freely in the region bounded between a square-shaped boundary of the box and the square-shaped scatterer. The particle reflects specularly from the boundary of the box and the scatterer in accordance with the Snell's law. }}
    \label{fig:square}
    \end{center}
\end{figure}

\begin{figure*}[t!]
    \centering
    \subfloat[]{\includegraphics[width=0.40\textwidth]{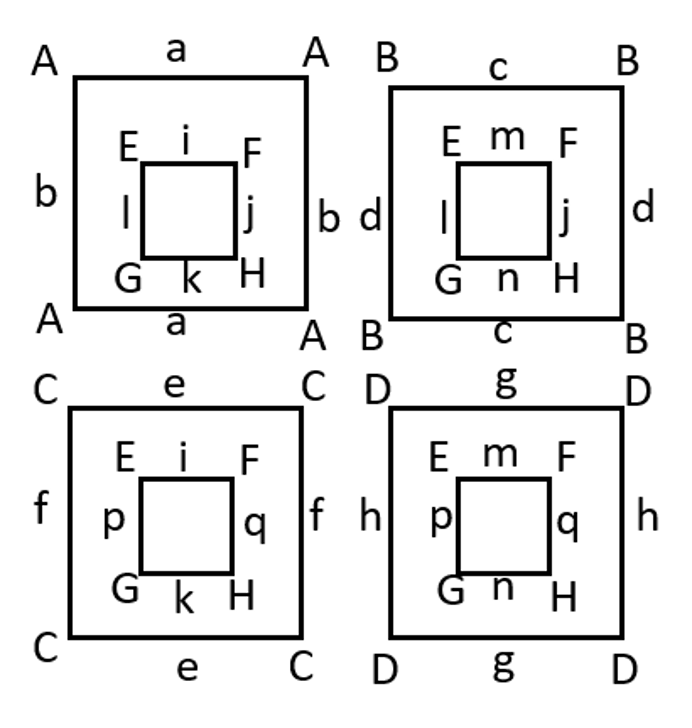}}\hspace{1mm}
    \subfloat[]{\includegraphics[width=0.42\textwidth]{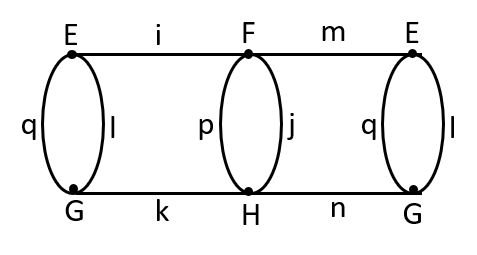}}\\ 
    \subfloat[]{\includegraphics[width=0.4\textwidth]{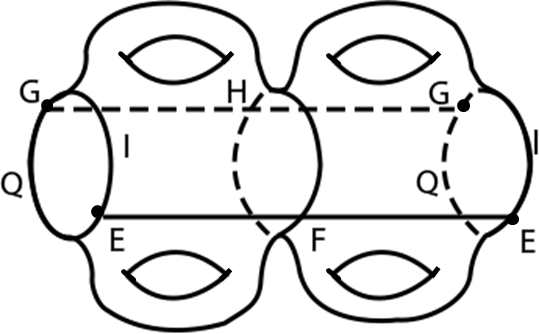}}\hspace{0.9mm}
    \subfloat[]{\includegraphics[width=0.3\textwidth]{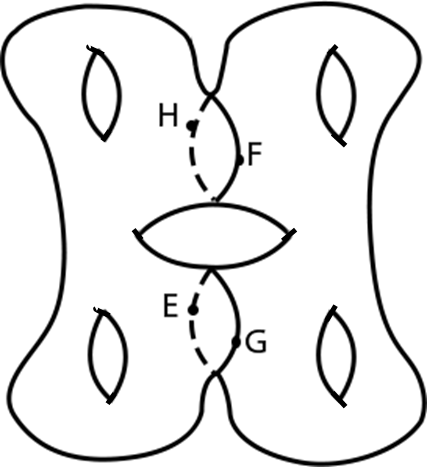}} 
    \caption{The fundamental domain consists of four copies of the square torus domain shown in (a). There are four squares with a square-shaped scatterer in each square with identified edges. In (b) and (c), construction of five handled sphere or a ``genus five surface" (d) is shown.}
\label{fig:G5}
\end{figure*}

The free motion of a point particle in a square torus billiard (STB) is shown in Figure \ref{fig:square}. According to the theorem by Zemlyakov and Katok \cite{zemlyakov}, this system is non-integrable albeit non-chaotic with zero Lyapunov exponent. The invariant integral surface is topologically equivalent to a sphere with five handles, as shown in \cite{richens1981pseudointegrable}. The entire trajectory of the free particle in the STB can be folded in four copies using which we can construct the invariant surface (constant energy). This is explained in Figure \ref{fig:G5}. In statistical mechanics, this model is related to  Ehrenfest gas where a beam of particles moving freely in a plane gets scattered by square-shaped scatterers (also called wind-tree model \cite{bob,manan}). A new finite-time exponent was introduced to describe these systems \cite{mcj} as the long-time average vanishes due to rather pathological behaviour of these systems. 

We shall now employ these features to our advantage in quantum encoding. 

\subsection{Encoding}
We start with the fundamental domain of an equivalent genus five surfaces, Fig. \ref{fig:G5}, obtained by tessellating a square with a square-shaped scatterer inside it four times and placing the data and the ancilla qubits alternatively on the vertex of external squares as well as on the vertex of scatterers. The data qubits are represented as $D$ (in the circles) and the ancilla qubits are represented as $A$ (in the squares). As in earlier sections, the bold (dashed) lines represent the control-$X(Z)$ operations from the ancilla qubits to the data qubits. The set of stabilizers is $P$=$\{X_1X_2X_3X_6X_7$, $X_3X_4X_5X_{12}X_{13}$, $X_1X_6X_8$, $X_2X_7X_9$, $X_3X_{10}X_{12}$, $X_3X_{11}X_{13}$, $Z_1Z_3Z_4Z_8Z_{10}$, $Z_2Z_3Z_5Z_9Z_{11}$, $Z_3Z_6Z_8$, $Z_3Z_7Z_9$, $Z_4Z_{10}Z_{12}$, $Z_5Z_{11}Z_{13}\}$. The logical state $\ket{0}_L$ is:
\begin{alignat}{1}\label{eq:zerol_g5}
    \ket{0}_L=&\frac{1}{\mathcal{N}}\prod_{P_i\in\langle P\rangle}(I^{\otimes n}+P_i)\ket{0^{\otimes n}}\nonumber\\
    =&\frac{1}{\mathcal{N}}(I^{\otimes 13}+X_1X_2X_3X_6X_7)(I^{\otimes 13}+X_3X_4X_5X_{12}X_{13})(I^{\otimes 13}+X_1X_6X_8)(I^{\otimes 13}+X_2X_7X_9)\nonumber\\
    &(I^{\otimes 13}+X_3X_{10}X_{12})(I^{\otimes 13}+X_3X_{11}X_{13})(I^{\otimes 13}+Z_1Z_3Z_4Z_8Z_{10})(I^{\otimes 13}+Z_2Z_3Z_5Z_9Z_{11})\nonumber\\
    &(I^{\otimes 13}+Z_3Z_6Z_8)(I^{\otimes 13}+Z_3Z_7Z_9)(I^{\otimes 13}+Z_4Z_{10}Z_{12})(I^{\otimes 13}+Z_5Z_{11}Z_{13})|0^{\otimes 13}\rangle .\nonumber\\ 
\end{alignat}

We next look for pairs of logical operators that commute with stabilizers and anti-commute pairwise. For this, we have to specify the boundaries. The filling of the plane using the fundamental domain of the equivalent genus five surfaces, forms periodically arranged branch cuts (edges $EF$ and $GH$ in Fig.\ref{fig:G5}), which are considered as the boundaries. Thus we define a path by connecting the data qubit vertex of one scatterer to the data qubit vertex of the corresponding copy with respect to the fundamental domain. The directed paths for the logical $\Bar{X}$ operator are: $X_6X_8X_{10}X_{12}$, $X_6X_8X_4X_{12}$, $X_7X_9X_{11}X_{13}$, and $X_7X_9X_{5}X_{13}$. The directed paths for the logical $\Bar{Z}$ operator are: $Z_8Z_6Z_7Z_9$, $Z_8Z_6Z_2Z_9$, $Z_8Z_1Z_7Z_9$, $Z_8Z_1Z_2Z_9$,  and $Z_{10}Z_{12}Z_{13}Z_{11}$. From these paths, we found a pair of logical operators $\{\Bar{X}=X_6X_8X_4X_{12}$, $\Bar{Z}=Z_8Z_1Z_7Z_9\}$. The minimum weight the error $E=E_a^\dagger E_b$, which violates the Knill-Laflamme conditions, is $3$, thereby constructing a $[[13,1,3]]$ code.
\begin{figure}
    \begin{center}
    \includegraphics[width=0.45\textwidth]{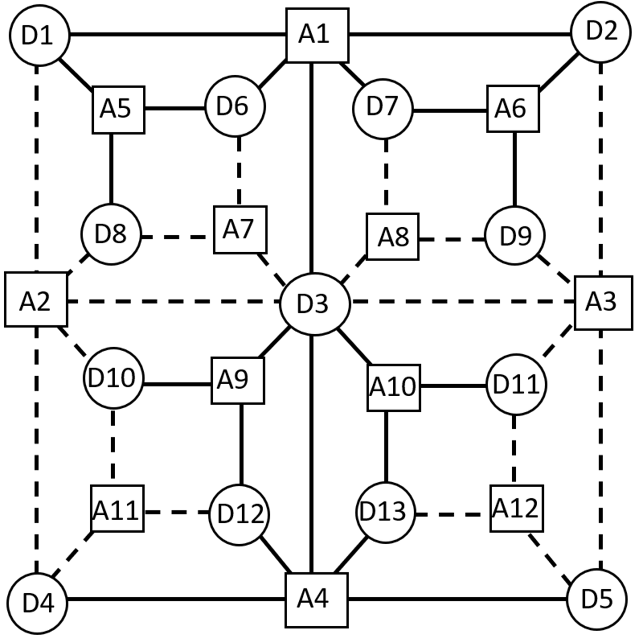}
    \caption{\small{A single unit of the code is constructed using the fundamental domain of equivalent genus five surface keeping the horizontal edges of the scatterer identical to the corresponding edge of other scatterer. The data qubits are represented by $D$ and ancillary qubits by $A$. The solid (dashed) lines represent $X$ ($Z$)-ancilla qubits. The data and ancillary qubits are placed alternately. The number of required data qubits is $n=13$ and the number of required ancilla qubits is $m=12$.}}
    \label{fig:g51_unit}
    \end{center}
\end{figure}

\begin{figure}
    \begin{center}
    \includegraphics[width=0.5\textwidth]{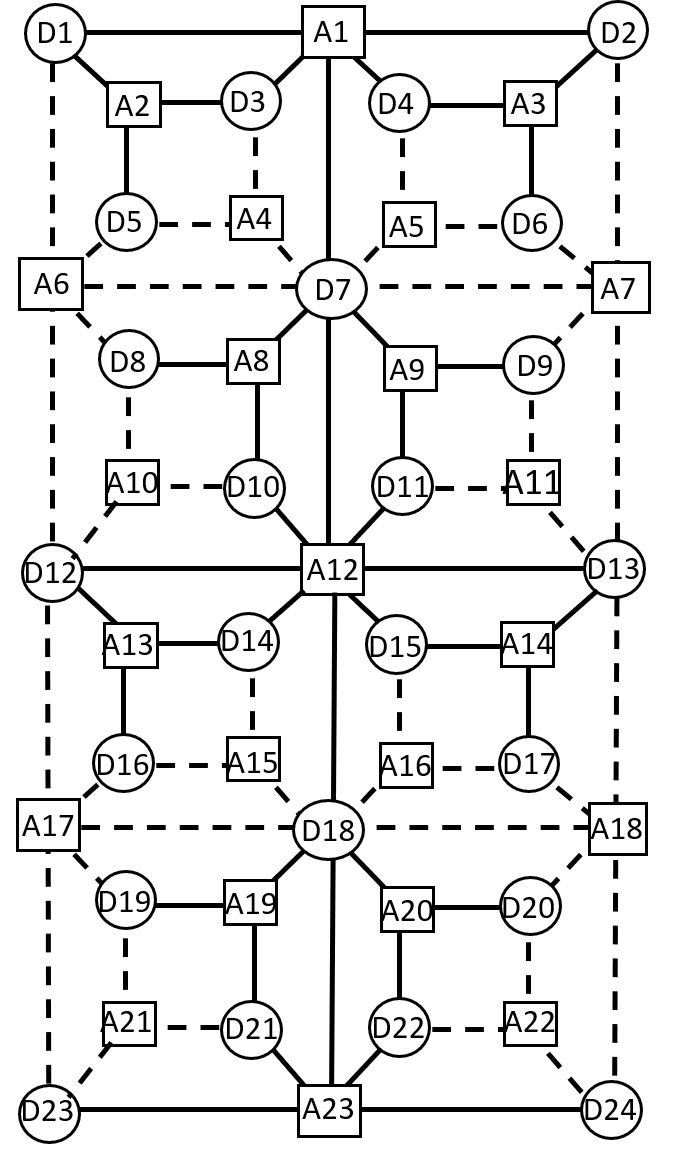}
    \caption{\small{The unit structure of the genus five code in Fig. \ref{fig:g51_unit} has been copied vertically. The number of data (ancilla) qubits required for this structure is $n=24$ ($m=23$). This increases the distance to $d=4$, constructing a $[[24,1,4]]$ code.}}
    \label{fig:g51_unit2}
    \end{center}
\end{figure}

To increase the distance of the code, we can stack the unit structure of the code (Fig. \ref{fig:g51_unit}) vertically as shown in Fig.\ref{fig:g51_unit2}. The number of required data qubits is $n=24$ and the number of required ancillary qubits is $m=23$. The set of
stabilizers is $P$=$\{X_1X_2X_3X_4X_7$, $X_1X_3X_5$, $X_2X_4X_6$, $X_7X_8X_{10}$, $X_7X_9X_{11}$, $X_7X_{10}X_{11}X_{12}X_{13}X_{14}X_{15}X_{18}$, $X_{12}X_{14}X_{16}$, $X_{13}X_{15}X_{17}$, $X_{18}X_{19}X_{21}$, $X_{18}X_{20}X_{22}$, $X_{18}X_{21}X_{22}X_{23}X_{24}$, $Z_3Z_5Z_7$, $Z_4Z_6Z_7$, $Z_1Z_5Z_7Z_8Z_{12}$, $Z_2Z_6Z_7Z_9Z_{13}$, $Z_8Z_{10}Z_{12}$, $Z_9Z_{11}Z_{13}$, $Z_{14}Z_{16}Z_{18}$, $Z_{15}Z_{17}Z_{18}$, $Z_{12}Z_{16}Z_{18}Z_{19}Z_{23}$, $Z_{13}Z_{17}Z_{18}Z_{20}Z_{24}$, $Z_{19}Z_{21}Z_{23}$, $Z_{20}Z_{22}Z_{24}\}$. The pair of logical operators is $\{\Bar{X}=X_8X_{12}X_{16}X_{14}$, $\Bar{Z}=Z_8Z_{10}Z_{15}Z_{17}\}$. The minimum weight that violates the Knill-Laflamme conditions for this code is $4$. Hence it is a $[[24,1,4]]$ code. Thus, the distance of the code can be increased by stacking fundamental domains on the plane.

\subsection{Effect of noise}
Any logical qubit should be robust against dephasing due to an external noise. Recently, it has been shown \cite{pal} that certain observables formed by code space population and logical operators in the code space help determine the dynamical behaviour of logical qubits. We incorporate a time-dependent external fluctuating magnetic field in $z$-direction, which  acts on the qubits globally, thus leading to global dephasing. To estimate the effect, consider the logical $|1\rangle_L$:
\begin{equation}
    |1\rangle_L= \Bar{X} \ket{0}_L
\end{equation}
Let an initial logical quantum state be written as
\begin{equation}
    \ket{\psi}_L=\cos{\frac{\theta}{2}}\ket{0}_L+e^{\iota\phi}\sin{\frac{\theta}{2}}\ket{1}_L
\end{equation}
where $\theta$ and $\phi$ are real parameters ($\theta\leq \pi$ and $0\leq \phi\leq 2\pi$). The evolution of $|\psi\rangle_L$ gives the logical Bloch sphere coordinates, $X_L$, $Y_L$ and $Z_L$. Assuming the global dephasing process by a single fluctuating variable $B(t)$ along the $z$-direction acting on all data qubits, the Hamiltonian representing the effect of noise may be written as $H_G(t) =\frac{1}{2}B(t)\sum_{i=1}^{13}\sigma_{z_i}$. In case of local dephasing, the Hamiltonian reads as: $H_L(t)=\frac{1}{2}\sum_{i=1}^{13}B_i(t) \sigma_{z_i}$. The randomly fluctuating variable $B(t)$ obeys the Gaussian distribution $P(B)$, which implies that \cite{pal}:
\begin{alignat}{1}
  \bigg\langle \exp&{\bigg(\pm \iota \int_{0}^{t}B(t^\prime)dt^\prime \bigg)}\bigg\rangle\nonumber\\
   &=\exp{\bigg[-\frac{1}{2}\bigg\langle\bigg(\int_{0}^{t}B(t^\prime) dt^\prime\bigg)^2\bigg\rangle}\bigg]= e^{-\gamma t/2}
\end{alignat}
assuming the stationarity of the auto-correlation function of delta-correlated noise, with  $\gamma=\braket{[B(0)]^2}$.

Following \cite{pal}, we analyze the effect of noise on the $N$-qubit system by grouping the physical states by their magnetization, defined as the difference between the number of spins in the state $\ket{0}$, denoted by $n^\prime$, and the remaining in state $\ket{1}$, $N-n^\prime$. The magnetisation is, $m^\prime=2n^\prime-N$. The logical state $\ket{0}_L$ is written as, $\ket{0}_L=\sum_{m^\prime} \sum_{l=1}^{N_{m^\prime}}b_l^{m^\prime}\ket{b}_l^{m^\prime}$. Dephasing noise changes the  state $\ket{\psi}_L$ to another state $\ket{\psi^\prime}$, where $\ket{\psi^\prime}=\exp{\bigg[-\iota\int_{0}^{t}H_{L, G}(t^\prime)dt^\prime\bigg]}\ket{\psi}_L$. The density matrix corresponding to the logical qubit is $\rho^\prime=\int \ket{\psi^\prime}\bra{\psi^\prime} P(B)dB.$
The Bloch coordinates $\mathcal{R}\equiv\{R_X, R_Y, R_Z\}$ in the new state are obtained by evaluating the expectation values of the logical operators in the evolved state, given by $\braket{{\mathcal{R}}}=Tr[\rho^\prime \bar{\mathcal{L}}]$, where $\bar{\mathcal{L}}\equiv\{\bar{X}, \bar{Y}, \bar{Z}\}$ represents the logical Bloch vectors in the initial state, $\ket{\psi}$. For the single unit structure (Fig. \ref{fig:g51_unit}), in the presence of global dephasing noise, the logical Bloch coordinates turn out to be
\begin{alignat}{1}
\braket{R_X}=&\frac{1}{32}e^{-(2\gamma t+\iota\phi)}(1+e^{-\gamma t})^4(1+e^{2\iota\phi})\sin{\theta}\nonumber\\
\braket{R_Y}=&\frac{\iota}{32}e^{-(2\gamma t+\iota\phi)}(1+e^{-\gamma t})^4(-1+e^{2\iota\phi})\sin{\theta}\nonumber\\
\braket{R_Z}=&\cos{\theta}\nonumber\\
\end{alignat}

In the absence of noise, i.e., $\gamma=0$, the Bloch sphere coordinates  in the new state, $\ket{\psi'}$ are $\braket{R_X}=\sin{\theta}\cos{\phi}$, $\braket{R_Y}=\sin{\theta}\sin{\phi}$, and $\braket{R_Z}=\cos{\theta}$ same as that in the old state, $\ket{\psi}_L$. Even in the presence of noise, $\langle R_Z\rangle$ remains unaffected. Thus the code is significantly robust against dephasing noise.

\section{Concluding remarks}
The basic idea underlying surface codes for error detection and correction is to be able to arrange the data and ancillary qubits in a way that $X$ and $Z$ errors can be corrected by making Stabilizer measurements through ancillae. For a scalable architecture, planar structures are desirable. This brings us to the question of tessellation of the plane. While in Kitaev's construction, two-dimensional Ising model is considered where the lattice shape can be anything - however, it should be noted that ``anything" is only under periodic boundary conditions where then, unit shapes could be square, equilateral triangle etc. Here we take the essence from Kitaev's construction and use the correspondence between Lie and reflection groups, ideas from well-known billiards, and present a novel way to realize architectures of higher genus. The encoding rates - number of logical qubits for the physical qubits - surpasses the value for all surface codes hitherto known. We believe that these results pave the way to a new direction of research in the field of quantum error correction.   

The codes presented here are not related to tessellations of hyperbolic surfaces. We have constructed fundamental domain using replicas of the billiard considered.  We then stack the domains, thus taking care of all the symmetries of the system. It is at this point that we endow each vertex with a qubit or ancilla. This enables us to write the stabilizers and construct logical operators. This construction respects the commutation and anticommutation relations expected of a consistent and complete definition of a code.

The spectra of the Hamiltonian made by the generators is studied. The degeneracy of the ground state increases with the number of qubits. For instance, for the genus-two codes $[[n, k, d]]$, the degeneracy of the ground state is $2^k$. The code is not topological. However, the ground state of the codes has high degeneracy which is useful for encoding. The code distance increases with the size of the code. The main advantage, however, is that the codes have much higher encoding rates. For genus-two codes of large size, the encoding rate tends to one-half. For the genus-five codes, the code distance increases with size whereas the encoding rate does not. Future investigations along these lines would be useful. 

In classical dynamical systems, tori as invariant surfaces are synonymous to integrability. The surfaces of higher genus correspond to non-integrability, but not chaos, even when the dynamics is nonlinear. Nonlinearity of the dynamics leads to the appearance of special points in the phase space, which have been shown to play an important role in controlling of quantum jumps for error correction \cite{krjj}. In quantum computing technology, almost all paradigms are related in an important way to aspects of nonlinearity, be it the nonlinearity of the Josephson junction, creation of EPR pair of photons from a nonlinear crystal and so on. Nonlinear resonances in coupled nonlinear quantum circuits with Josephson junctions have been shown to provide criteria for protection of qubits \cite{ssj}. Ideas from nonlinear science would expectedly contribute to the development of quantum information theory and technology. 

\vskip 0.25 truecm
\noindent
{\bf Acknowledgements}\\
Authors thank the Referee for her(his) critique drawn on our work. They also thank Rhine Samajdar, Princeton University, for several helpful and stimulating discussions. 

\vskip 0.25 truecm
\noindent
{\bf Data Availability Statement}: No Data associated in the manuscript


\begin{thebibliography}{99}
\bibitem{kvant} Ed. S. Tabachnikov, {\em Kvant Selecta: Algebra and Analysis, I and II} (Universities Press (India) Limited, 2002).  

\bibitem{weissman} M. H. Weissman, {\em An illustrated theory of numbers} (American Mathematical Society, 2017).

\bibitem{aop2014} R. Samajdar and S. R. Jain, Ann. Phys. {\bf 351}, 1 (2014).

\bibitem{aop2016} N. Manjunath, R. Samajdar, S. R. Jain, Ann. Phys. {\bf 372}, 68 (2016).

\bibitem{rmp2017} S. R. Jain and R. Samajdar, Rev. Mod. Phys. {\bf 89}, 045005 (2017). 

\bibitem{nakahara} M. Nakahara, {\em Geometry, Topology, and Physics} (Taylor and Francis, London, 2003).

\bibitem{coxeter} H. S. M. Coxeter, {\em Regular Polytopes} (Dover, New York, 1973). 

\bibitem{toffoli} E. Fredkin and T. Toffoli (1982), International Journal of Theoretical Physics {\bf 21}, 219 (1982). 

\bibitem{krj} K. Kumari, G. Rajpoot, and S. R. Jain, {\em A genus-two surface code} (arXiv:2211.12695 [quant-ph]).

\bibitem{weyl1926nachtrag} Hermann Weyl, Mathematische Zeitschrift {\bf 24}, 789 (1926).

\bibitem{cartan1927geometrie} {\'E}lie Cartan, Annali di Matematica pura ed applicata {\bf 4}, 209 (1927).

\bibitem{arnold} V. I. Arnol'd, {Mathematical methods of classical mechanics} (Springer, Heidelberg, 1978). 

\bibitem{jain1992} S. R. Jain and H. D. Parab, J. Phys. A{\bf 25}, 6669 (1992). 

\bibitem{Kitaev} Alexei Kitaev, Ann. Phys. {\bf 303}, 2 (2003).

\bibitem{eckhardt1984analytically} Bruno Eckhardt, Joseph Ford and Franco Vivaldi, Physica D: Nonlinear Phenomena {\bf 13}, 339--356 (1984).

\bibitem{zemlyakov} A. Zemlyakov and A. B. Katok, Math. Notes {\bf 18}, 760 (1976).

\bibitem{richens1981pseudointegrable} P. J. Richens and M. V. Berry, Physica D: Nonlinear Phenomena {\bf 2}, 495--512 (1981).

\bibitem{Gottesman} Daniel Gottesman, {\em Stabilizer codes and quantum error correction}, Ph. D. thesis (California Institute of Technology, 1997).

\bibitem{aa} V. I. Arnold and A. Avez, {\em Ergodic problems of classical mechanics} (W. A. Benjamin, Inc., Amsterdam, 1970). 

\bibitem{bob} J. R. Dorfman, {An introduction to chaos in nonequilibrium statistical mechanics} (Cambridge Univ. Press, Cambridge, 1999).

\bibitem{manan} M. Jain, Student J. Phys. {\bf 5}, 55 (2013). 

\bibitem{mcj} S. Moudgalya, S. Chandra, and S. R. Jain, Ann. Phys. {\bf 361}, 82 (2015).

\bibitem{pal} Amit Kumar Pal, Philipp Schindler, Alexander Erhard, {\'A}ngel Rivas, Miguel A. Martin-Delgado, Rainer Blatt, Thomas Monz and Markus P. M{\"u}ller, Quantum {\bf 6}, 632 (2022).

\bibitem{krjj} K. Kumari, G. Rajpoot, S. Joshi, and S. R. Jain, Ann. Phys. {\bf 450}, 169222 (2023).

\bibitem{ssj} R. K. Saini, R. Sehgal, and S. R. Jain, Eur. Phys. J. Plus {\bf 137}, 356 (2022).

\end{thebibliography}
\end{document}